\documentclass[11pt]{cernrep}
\usepackage{here}
\usepackage{amsbsy}
\usepackage{amsfonts}
\usepackage{amssymb}
\usepackage[dvips]{graphicx}
\begin{document}
\title{The Power of Confidence Intervals}
\author{Carlo Giunti$^{\mathrm{(a)}}$ and Marco Laveder$^{\mathrm{(b)}}$}
\institute{$^{\mathrm{(a)}}$
INFN, Sezione di Torino,
and
Dipartimento di Fisica Teorica,
Universit\`a di Torino,
Via P. Giuria 1, I--10125 Torino, Italy
\\
$^{\mathrm{(b)}}$
Dipartimento di Fisica ``G. Galilei'', Universit\`a di Padova,
and INFN, Sezione di Padova,
Via F. Marzolo 8, I--35131 Padova, Italy}
\maketitle
\begin{abstract}
We connect the power
of Confidence Intervals
in different Frequentist methods
to their reliability.
We show that
in the case of a bounded parameter
a biased method which
near the boundary has
large power in testing the parameter against larger alternatives
and
small power in testing the parameter against smaller alternatives
is desirable.
Considering the recently proposed methods with correct coverage,
we show that
the Maximum Likelihood Estimator method \cite{Ciampolillo-98,Mandelkern-Schultz-99}
has optimal bias.
\end{abstract}

It is well known that the most important
property of Frequentist Confidence Intervals
is \emph{coverage}:
a $100(1-\alpha)\%$ Confidence Interval
belong to a set of intervals that cover the true value of the
measured quantity $\mu$ with
Frequentist probability $1-\alpha$.
Neyman's method obtains Confidence Intervals
with correct coverage
through the construction
for each possible value of $\mu$ of an
\emph{acceptance interval}
with probability $1-\alpha$
for an estimator $\widehat{\mu}$ of $\mu$.
The union of all
acceptance intervals
in the $\widehat{\mu}$--$\mu$
plane
is called the \emph{Confidence Belt}.
The Confidence Interval for $\mu$
resulting from a measurement $\widehat{\mu}_{\mathrm{obs}}$
of the estimator is the set of all values of $\mu$
whose acceptance interval for $\widehat{\mu}$ include
$\widehat{\mu}_{\mathrm{obs}}$.

Coverage is not the only property of Confidence Intervals,
because many methods for the construction of a Confidence Belt with exact coverage
are available
(see Refs.~\cite{Kendall-2A,Feldman-Cousins-98,Giunti-bo-99,Ciampolillo-98,Mandelkern-Schultz-99}).
These methods differ by \emph{power}
\cite{Giunti:2000kc},
a quantity which is obtained considering the
construction of acceptance intervals as hypothesis testing.
Coverage and power are connected, respectively, with the so-called
\emph{Type I} and \emph{Type II} errors in testing a
simple statistical hypothesis $H_0$
against a simple alternative hypothesis $H_1$
(see Ref.~\cite{Kendall-2A}, section 20.9):
\begin{description}
\item[Type I error:]
Reject the null hypothesis $H_0$ when it is true.
The probability of a Type I error is called
\emph{size} of the test and it is usually denoted by $\alpha$.
\item[Type II error:]
Accept the null hypothesis $H_0$ when the alternative hypothesis $H_1$ is true.
The probability of a Type II error is usually denoted by $\beta$.
The power of a test is the probability $\pi=1-\beta$
to reject $H_0$ if $H_1$ is true.
A test is \emph{Most Powerful}
if its power
is the largest one among all possible tests.
This is clearly the best choice.
\end{description}

Unfortunately,
the power associated with a confidence belt
is not easy to evaluate,
because for each possible value $\mu_0$ of $\mu$
considered as a null hypothesis
there is no simple alternative hypothesis
that allows to calculate the probability $\beta$
of a Type II error.
Instead,
we have the alternative hypothesis
$H_1$: $\mu_1\neq\mu_0$,
which is composite.
For each value of $\mu_1\neq\mu_0$
one can calculate the probability $\beta_{\mu_0}(\mu_1)$
of a Type II error associated with a given acceptance interval
corresponding to $\mu_0$.
A method that gives an acceptance region
for $\mu_0$
which has the largest possible power
$\pi_{\mu_0}(\mu_1)=1-\beta_{\mu_0}(\mu_1)$
is Most Powerful with respect to the alternative $\mu_1$.
Clearly,
it would be desirable to find a
\emph{Uniformly Most Powerful}
test,
\textit{i.e.}
a test that gives an acceptance region
for $\mu_0$
which has the largest possible power
$\pi_{\mu_0}(\mu_1)$
for any value of $\mu_1$.
Unfortunately,
the Neyman-Pearson lemma implies that
in general a Uniformly Most Powerful test
does not exist if the alternative hypothesis is
\emph{two-sided},
\textit{i.e.}
both $\mu_1<\mu_0$ and $\mu_1>\mu_0$ are possible,
and the derivative of the Likelihood with respect to $\mu$
is continuous in $\mu_0$
(see Ref.~\cite{Kendall-2A}, section 20.18).
Nevertheless,
it is possible to find a Uniformly Most Powerful test
if the class of tests is restricted in appropriate ways.
A class of tests that has some merit
is that of \emph{unbiased} tests,
such that
the power $\pi_{\mu_0}(\mu_1)$
for any value of $\mu_1$
is larger or equal to the size $\alpha$ of the test,
\begin{equation}
\pi_{\mu_0}(\mu_1)
\geq
\alpha
\qquad
\mbox{for all $\mu_1$}
\,.
\label{unbiased}
\end{equation}
In other words,
the probability of rejecting $\mu_0$ when it is false
is at least as large as the probability of
rejecting $\mu_0$ when it is true.
The \emph{equal-tail} test
used in the Central Intervals method
is unbiased
and \emph{Uniformly Most Powerful Unbiased}
for distributions belonging to the exponential family,
such as, for example,
the Gaussian and Poisson distributions
(see Ref.~\cite{Kendall-2A}, section 21.31).

Therefore,
the Central Intervals method
is widely used because it corresponds to
a Uniformly Most Powerful Unbiased test.
Other methods based on asymmetric tests unavoidably introduce
some bias.

\begin{figure}
\begin{center}
\mbox{
\includegraphics[bb=173 415 502 677, width=0.49\textwidth, clip]{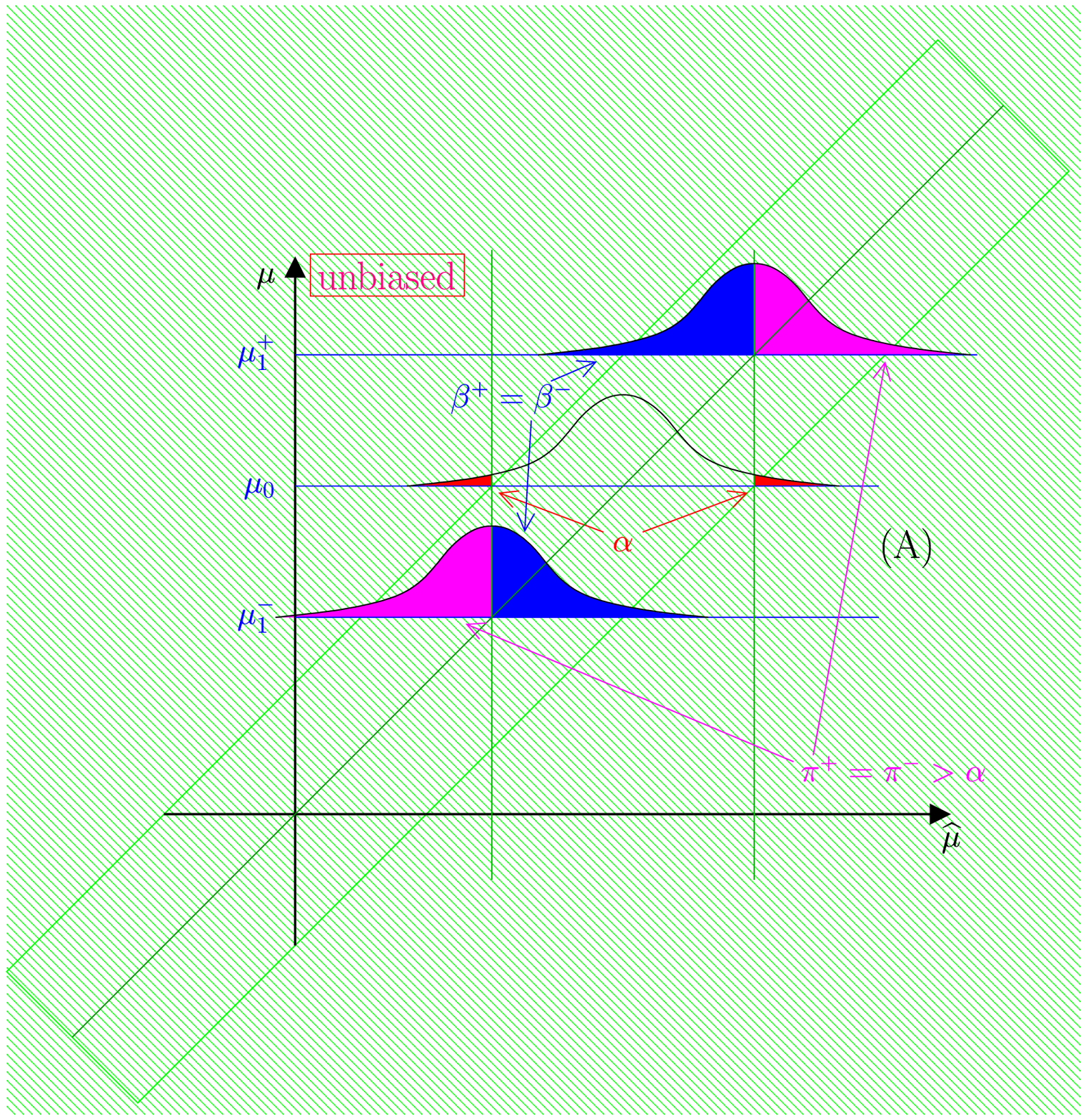}
\includegraphics[bb=100 399 500 700, width=0.49\textwidth, clip]{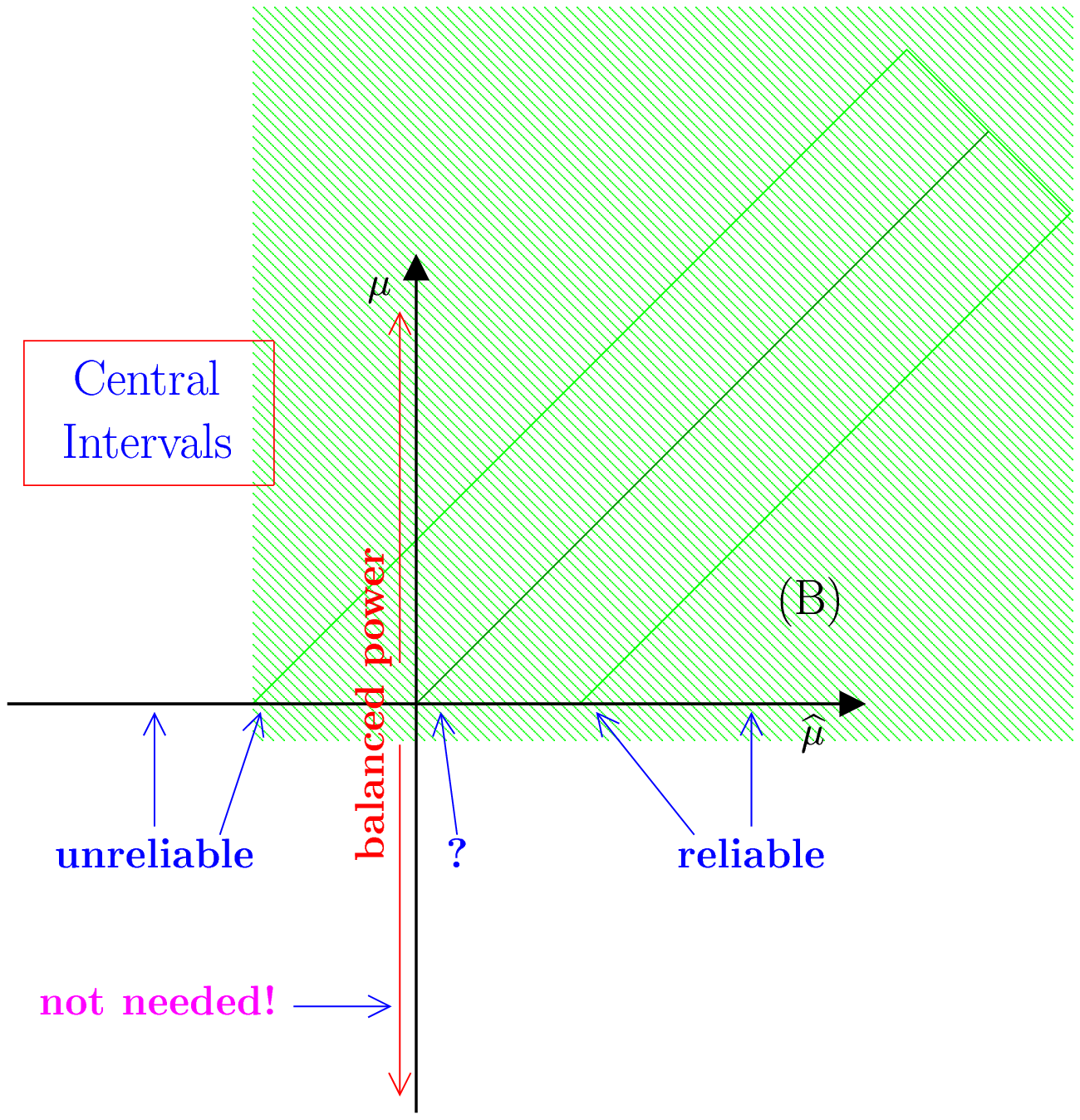}
}
\end{center}
\caption{ \label{ci}
A: Power $\pi$ in the Central Intervals method
for an estimator $\widehat{\mu}$ of $\mu$
that has a Gaussian distribution.
B: Reliability of the Confidence Intervals obtained
with the Central Intervals method
for a bounded $\mu\geq0$.
See text for details.
}
\end{figure}

Figure~\ref{ci}A
illustrates the power $\pi$ in the Central Intervals method
for an estimator $\widehat{\mu}$ of $\mu$
that has a Gaussian distribution.
The Gaussian distribution
of $\widehat{\mu}$ for $\mu=\mu_0$
is depicted qualitatively above the horizontal line
for $\mu=\mu_0$.
The $100(1-\alpha)\%$ acceptance interval corresponding to 
the null hypothesis $\mu_0$
is limited by the two vertical lines.
The area of the two dark-shaded tails of the distribution
is equal to $\alpha$.

Let us consider for example
the alternative hypothesis $\mu_1^+>\mu_0$
(similar considerations apply to the alternative hypothesis $\mu_1^-<\mu_0$).
The Gaussian distribution
of $\widehat{\mu}$ for $\mu=\mu_1^+$
is depicted qualitatively above the horizontal line
for $\mu=\mu_1^+$ in Fig.~\ref{ci}A.
The probability $\beta^+$
of a Type II error in testing $\mu_0$ against $\mu_1^+$
is given by the integral of the distribution
of $\widehat{\mu}$ for $\mu=\mu_1^+$
in the interval between the two horizontal lines.
The corresponding area is shown dark-shaded in Fig.~\ref{ci}A.
The power
to test the null hypothesis $\mu_0$
against the larger alternative $\mu_1^+>\mu_0$,
is given by the integral of the distribution of
$\widehat{\mu}$
for $\mu=\mu_1^+$
in the two semi-infinite intervals of $\widehat{\mu}$
external to the two horizontal lines.
The corresponding areas are shown light-shaded in Fig.~\ref{ci}A
(only the one on the right is large enough to be visible).

From Fig.~\ref{ci}A
one can see that the power corresponding to
alternative hypotheses $\mu_1^-$ and $\mu_1^+$,
respectively smaller and larger than the null hypothesis
$\mu_0$,
is equal.
The Central Intervals method
produces the most reliable results
in the case of an unbounded $\mu$,
because the power is perfectly balanced.
Problems arise if one considers the measurement of a bounded quantity $\mu$.
As illustrated in Fig.~\ref{ci}B
for the case of a bounded $\mu\geq0$,
the balanced power in the Central Intervals method
is not appropriate.
Indeed,
a high power to test $\mu_0$ against $\mu_1^-<\mu_0$
when $\mu_0$ is near the boundary
is not needed, because the alternatives
$\mu_1^-<\mu_0$
are limited.
As a result,
the Central Intervals method
produces in this case
clearly unreliable Confidence Intervals
if the value of
$\widehat{\mu}_{\mathrm{obs}}$
lies on the left-hand side of Fig.~\ref{ci}B.
Sometimes the Confidence Interval can be empty,
giving no information.
Sometimes one can get a very stringent upper limit,
much smaller than the exclusion potential of the experiment
\cite{Feldman-Cousins-98,Giunti:2000cd}.
This possibility
is very dangerous,
because it can lead to wrong conclusions
if interpreted in inappropriate ways.
In any case it gives no useful information on the value of $\mu$.

\begin{figure}
\begin{center}
\mbox{
\includegraphics[bb=144 398 496 753, width=0.45\textwidth, clip]{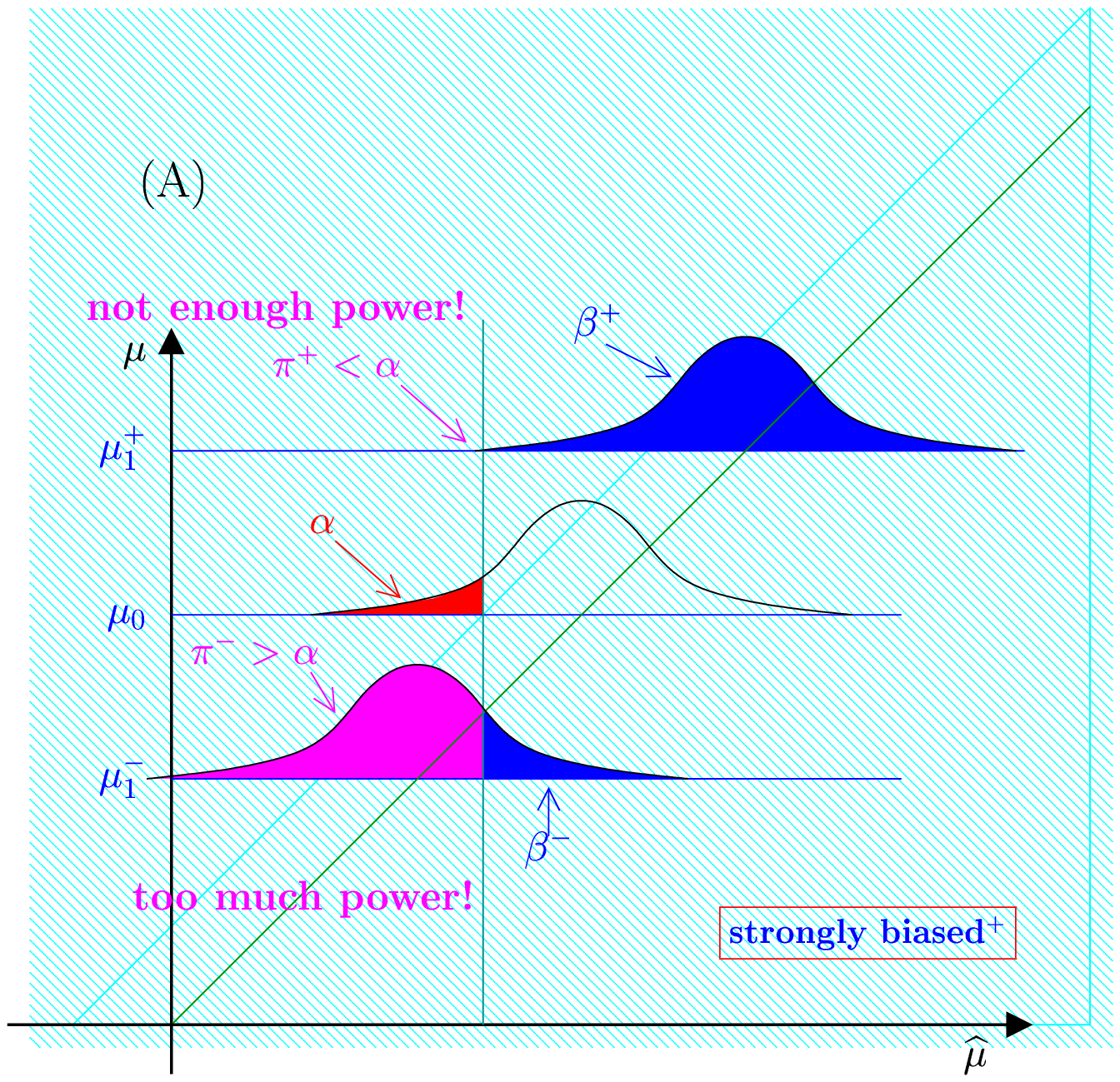}
\includegraphics[bb=128 401 426 706, width=0.45\textwidth, clip]{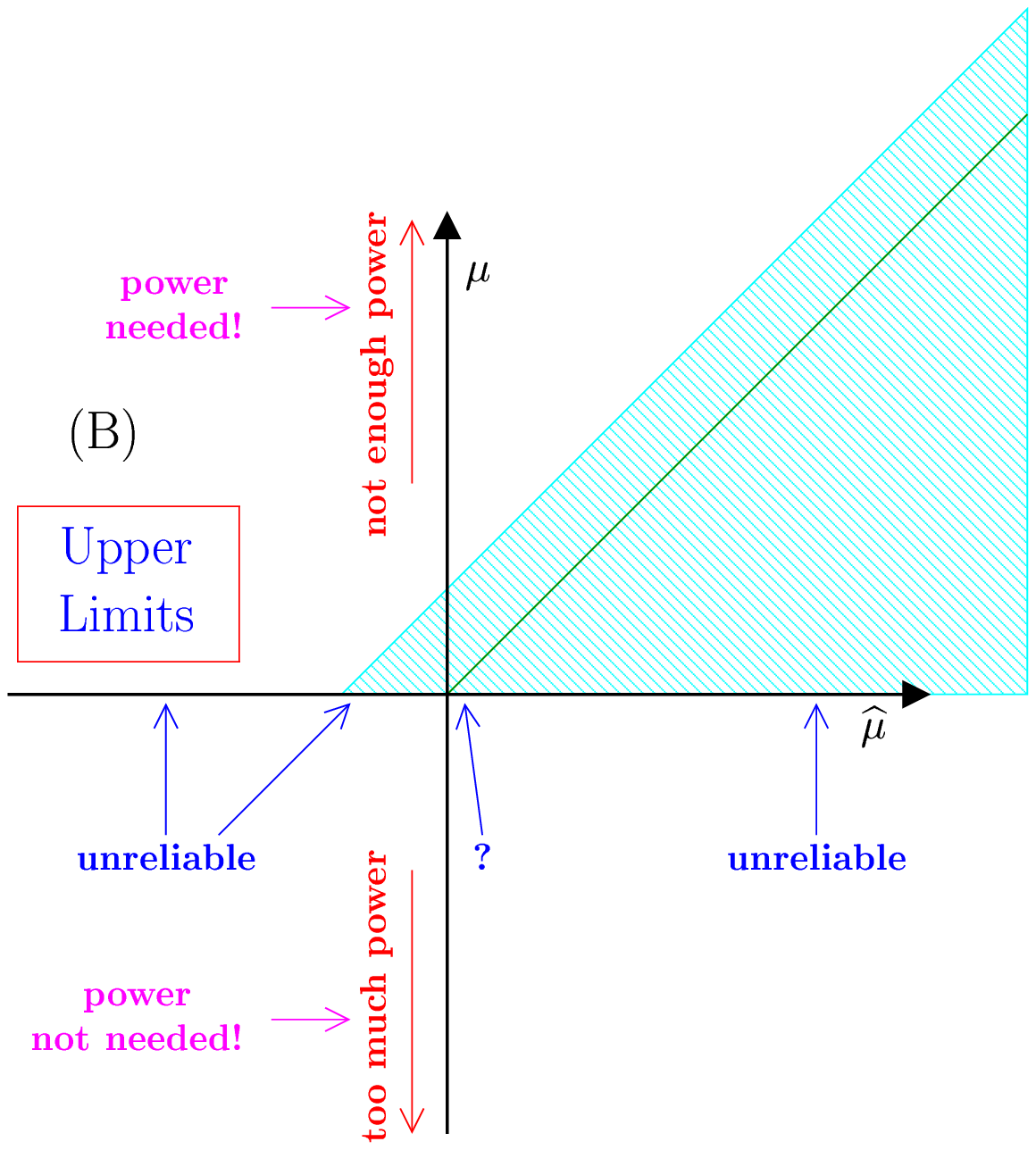}
}
\end{center}
\caption{ \label{ul}
A: Power $\pi$ in the Upper Limits method
for an estimator $\widehat{\mu}$ of $\mu\geq0$
that has a Gaussian distribution.
B: Reliability.
}
\end{figure}

In the past the Upper Limits method was rather popular.
Figures~\ref{ul}A and \ref{ul}B
show that the Upper Limits method is actually worse
than the Central Intervals method
because it is biased in the wrong direction.
As a consequence,
it produces limits that are practically always unreliable,
except maybe when by chance $\widehat{\mu}_{\mathrm{obs}} \simeq 0$.

\begin{figure}
\begin{center}
\mbox{
\includegraphics[bb=187 545 507 678, width=0.49\textwidth, clip]{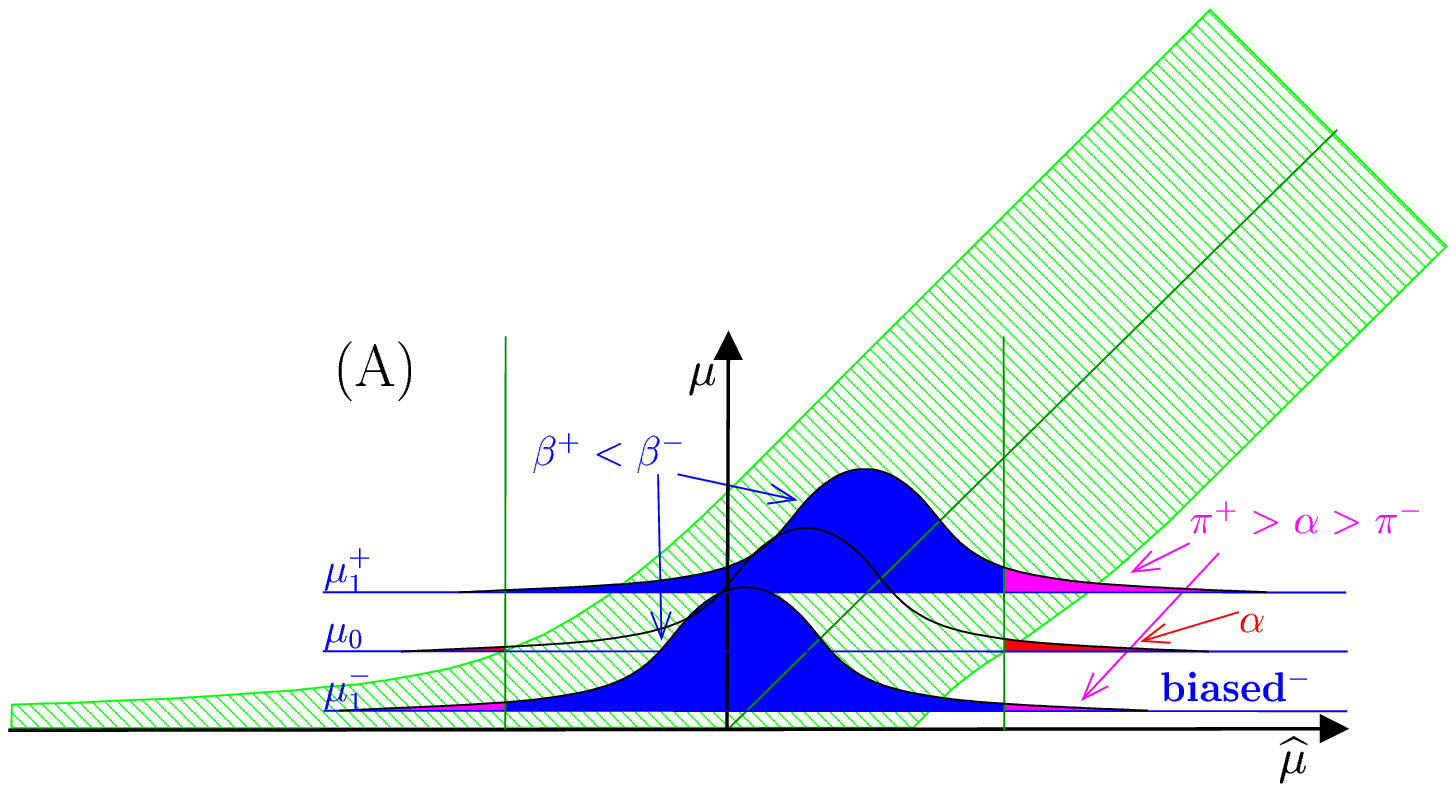}
\includegraphics[bb=111 506 479 685, width=0.49\textwidth, clip]{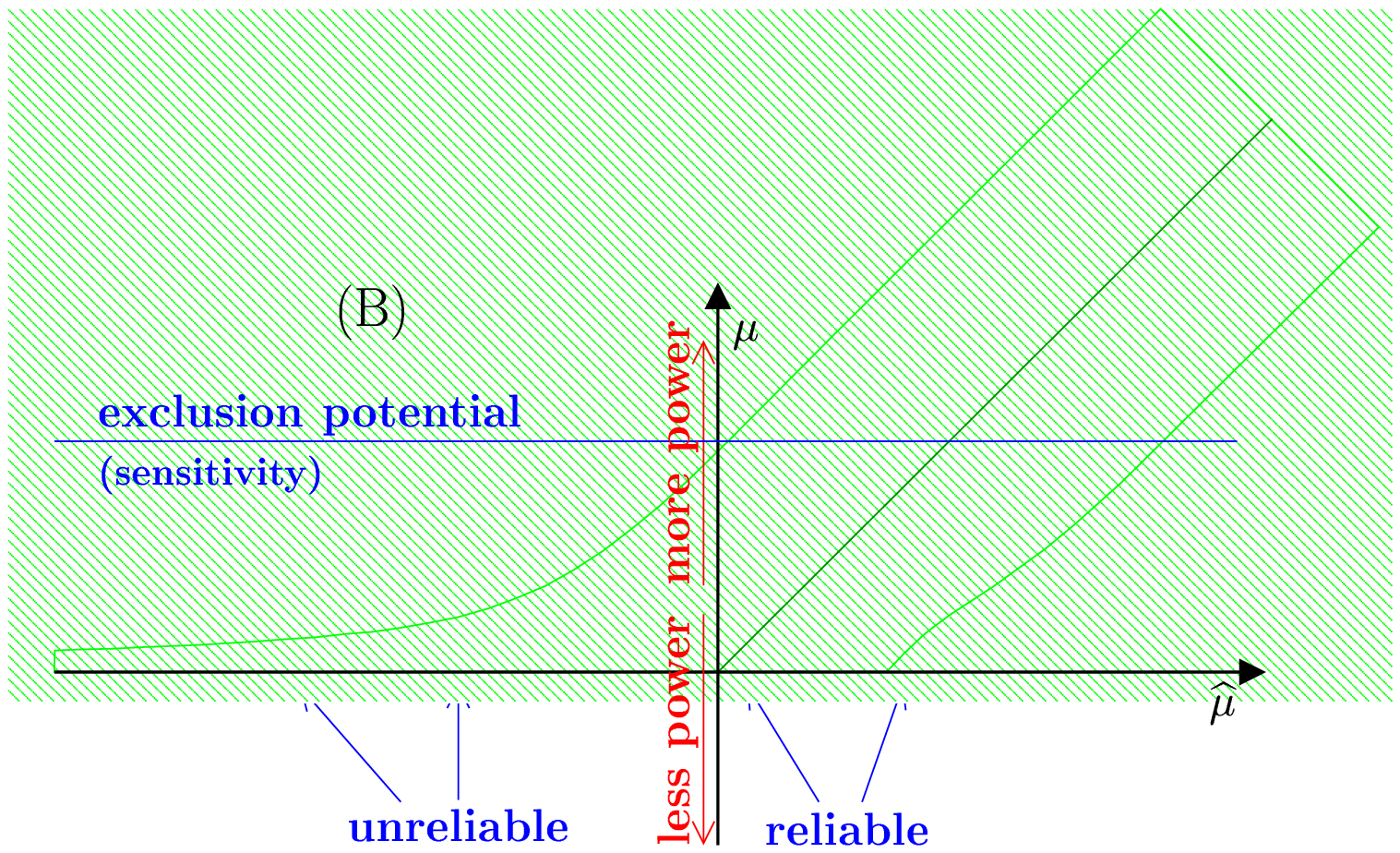}
}
\end{center}
\caption{ \label{ua}
A: Power $\pi$ in the Unified Approach
for an estimator $\widehat{\mu}$ of $\mu\geq0$
that has a Gaussian distribution.
B: Reliability.
}
\end{figure}

The method biased in the right direction
that has been proposed first
is the Unified Approach of Feldman and Cousins
\cite{Feldman-Cousins-98},
which,
as illustrated in Fig.~\ref{ua}A,
gives more power to test
$\mu_0$ against $\mu_1^+>\mu_0$
than to test
$\mu_0$ against $\mu_1^-<\mu_0$
when $\mu_0$ is near the boundary.
However,
the bias is still insufficient to produce reliable results
if $\widehat{\mu}_{\mathrm{obs}} \ll 0$:
from Fig.~\ref{ua}B one can see that when
$\widehat{\mu}_{\mathrm{obs}} \ll 0$
the Confidence Interval gives an upper limit for $\mu$
that is unphysically too small
\cite{Giunti-bo-99,Roe:1998zi,Mandelkern-Schultz-99,%
Giunti-CERN-2000-005,Astone-CERN-2000-005},
much smaller than the exclusion potential of the experiment
\cite{Feldman-Cousins-98,Giunti:2000cd}.

\begin{figure}
\begin{center}
\mbox{
\includegraphics[bb=156 545 463 679, width=0.49\textwidth, clip]{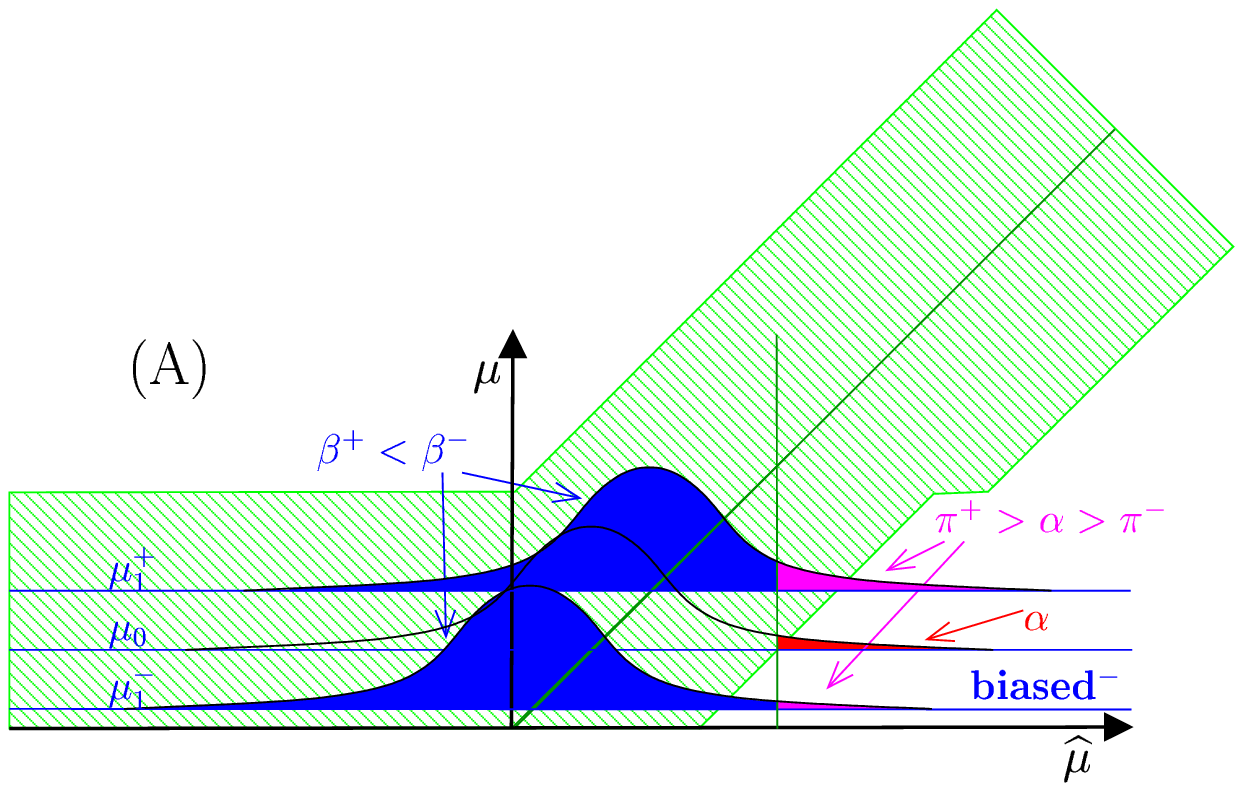}
\includegraphics[bb=95 420 489 641, width=0.49\textwidth, clip]{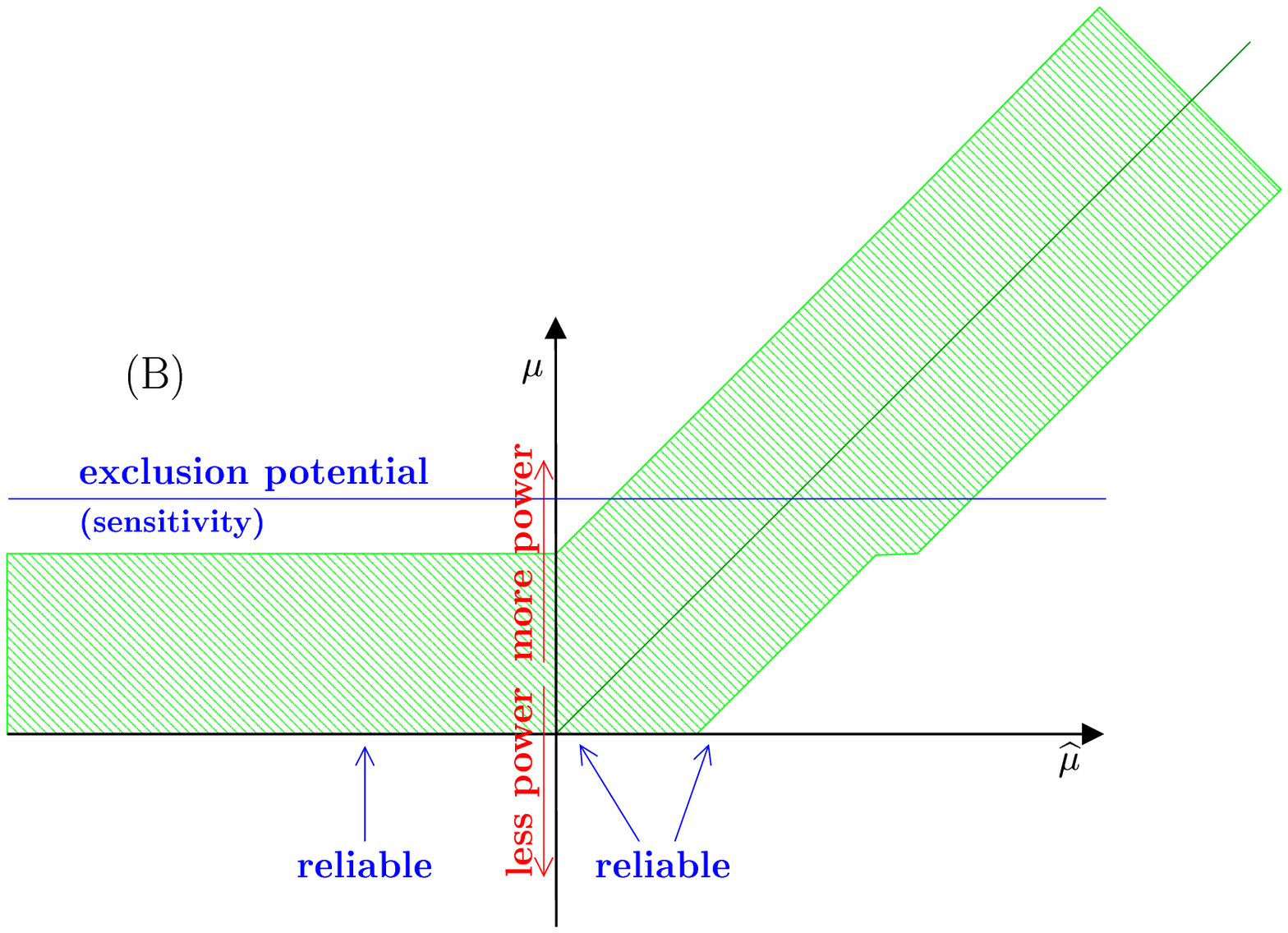}
}
\end{center}
\caption{ \label{ml}
A: Power $\pi$ in the Maximum Likelihood method
for an estimator $\widehat{\mu}$ of $\mu\geq0$
that has a Gaussian distribution.
B: Reliability.
}
\end{figure}

Figure~\ref{ml}A
illustrates the calculation of the power in the
Maximum Likelihood Estimator method proposed independently by
Ciampolillo in Ref.~\cite{Ciampolillo-98}
and
Mandelkern and Schultz in Ref.~\cite{Mandelkern-Schultz-99}.
In this method the estimator of $\mu$ is not $\widehat{\mu}$,
but the maximum likelihood value  $\mu^*$ of $\mu$.
Since the range of $\mu^*$ is equal to the range of $\mu$,
the estimate $\mu^*_{\mathrm{obs}}$
always lies in the physical range of $\mu$.
In the case of a Gaussian distribution for $\widehat{\mu}$
illustrated in Fig.~\ref{ml}A,
$\mu^*=\widehat{\mu}$
for $\widehat{\mu} \geq 0$
and
$\mu^*=0$
for $\widehat{\mu} \leq 0$.
Therefore,
as shown in Fig.~\ref{ml}A,
the upper limit for $\mu$
obtained for any $\widehat{\mu}_{\mathrm{obs}} < 0$
is equal to the upper limit obtained for $\widehat{\mu}_{\mathrm{obs}} = 0$.

As one can see from Fig.~\ref{ml}A,
the Maximum Likelihood Estimator method has optimal bias.
As a consequence,
this method
produces reliable results for any value of
$\widehat{\mu}_{\mathrm{obs}}$,
as shown in Fig.~\ref{ml}B.

Let us emphasize that the bias
is needed near the boundary and both the
Maximum Likelihood Estimator method
and
the 
Unified Approach
produce Confidence Intervals
that practically coincide with those obtained with
the Central Intervals method
when
$\widehat{\mu}_{\mathrm{obs}} \gg 0$.

In conclusion,
we have shown that the Maximum Likelihood Estimator method
\cite{Ciampolillo-98,Mandelkern-Schultz-99}
have optimal power in the case
of measurement of a bounded quantity
and produces always reliable
Confidence Intervals.
For these reasons,
it should be preferred over the
Unified Approach
\cite{Feldman-Cousins-98},
which is however better than the Central Intervals method.
Worse of all is the method of Upper Limits.

\end{document}